\documentclass[reprint,amsmath,amssymb,aps,prc]{revtex4-2}

\usepackage{graphicx}
\usepackage{dcolumn}
\usepackage{bm}
\usepackage{epstopdf}
\usepackage[mathlines]{lineno}
\usepackage{harpoon}
\usepackage{makecell}
\usepackage{subfigure}
\usepackage{hyperref}
\usepackage{color}
\usepackage{braket}
\usepackage{amsmath}
\usepackage{mathrsfs}
\usepackage{appendix}
\usepackage{multirow}
\usepackage{float}
\usepackage{enumerate}
\usepackage{ulem}
\usepackage{slashed}
\usepackage{orcidlink}

\begin{document}

\title{Examination of the evidence for a proton halo in $^{22}$Al}

\author{K. Y. Zhang\orcidlink{0000-0002-8404-2528}}
\affiliation{Institute of Nuclear Physics and Chemistry, China Academy of Engineering Physics, Mianyang, Sichuan 621900, China}

\author{C. Pan\orcidlink{0000-0003-3675-8238}} \email{cpan@ahnu.edu.cn}
\affiliation{Department of Physics, Anhui Normal University, Wuhu, Anhui 241000, China}

\author{Sibo Wang\orcidlink{0000-0002-2050-0040}}
\affiliation{Department of Physics and Chongqing Key Laboratory for Strongly Coupled Physics, Chongqing University, Chongqing 401331, China}

\date{\today}

\begin{abstract}
  More and more halo nuclei or candidates have been identified or suggested in experiments in recent years. It was declared that the halo structure of $^{22}$Al is revealed by the large isospin asymmetry in $^{22}$Si/$^{22}$O mirror Gamow-Teller transitions [\href{https://doi.org/10.1103/PhysRevLett.125.192503}{Phys. Rev. Lett. 125, 192503 (2020)}]. We highlight that a significant mirror asymmetry already exists between wave functions of the likely unbound nucleus $^{22}$Si and the doubly-magic nucleus $^{22}$O, which largely explains the observed asymmetry in the transitions. Furthermore, these transitions involve only the $1^+$ excited states of the daughter nuclei $^{22}$Al and $^{22}$F. The $1^+$ state of $^{22}$Al cannot be considered a halo state due to its proton-unbound nature. An analysis of the spin parity suggests that a weakly bound $2s_{1/2}$ valence proton in the ground state of $^{22}$Al is improbable. To investigate the shell structure for the ground state of $^{22}$Al, we employ the state-of-the-art deformed and triaxial relativistic Hartree-Bogoliubov theories in continuum. We find that a small $s$-wave component of $5\%$ appears for the weakly bound valence proton in $^{22}$Al only when triaxial deformation is considered. While the examination of densities and rms radii indicates that this small $s$-wave component is insufficient to form a discernible proton halo in $^{22}$Al, slightly larger $2s_{1/2}$ occupations have been reported in other recent theoretical results. The question of how many low-$\ell$ components are sufficient to form a proton halo in the presence of the Coulomb barrier remains open. Thus, future measurements of reaction or interaction cross sections and momentum distributions of breakup fragments are highly desirable to verify whether $^{22}$Al is a halo nucleus.
\end{abstract}

\date{\today}

\maketitle

\section{Introduction}

In 1985, Tanihata \textit{et al}. \cite{Tanihata1985PRL} measured the interaction cross sections $\sigma_I$ for $^{6\text{--}9,11}$Li on targets Be, C, and Al at 790 MeV/nucleon.
The root-mean-square (rms) radii of these lithium isotopes were deduced from the measured $\sigma_I$ by a Glauber-type calculation.
Surprisingly, $^{11}$Li exhibited a remarkably larger radius, $3.27\pm0.24$ fm, compared with those of approximately 2.5 fm for $^{6\text{--}9}$Li.
This was attributed to a large deformation or a long tail in the matter distribution in $^{11}$Li.

In 1988, Kobayashi \textit{et al}. \cite{Kobayashi1988PRL} measured the projectile fragmentation of $^{11}$Li at 790 MeV/nucleon.
An extremely narrow width of the transverse momentum distribution of the $^{9}$Li fragmentation was observed.
This narrow width was understood to originate from the removal of the two weakly bound outer neutrons in $^{11}$Li.
They form a halo-like structure enveloping the core nucleus $^{9}$Li, and such an unprecedented, exotic phenomenon is thus termed ``neutron halo."

Since then, it has become a paradigm that the identification of a halo nucleus necessitates the observation of both an enhancement in the cross section and a narrow momentum distribution of the breakup fragments \cite{Tanihata2013PPNP}.
In addition, some other features such as a low nucleon separation energy and a soft $E1$ excitation are indicative of a possible halo \cite{Tanihata2023Book,Aumann2013PhysScr}.

To date, about 20 halo nuclei or candidates have been identified or suggested in experiments (see Fig. 1.4 of Ref. \cite{Tanihata2013PPNP} and Fig. 1 of Ref. \cite{Zhang2023PRC(L1)}).
It is noteworthy that there also could be proton halos in proton-rich nuclei, e.g., $^8$B \cite{Warner1995PRC(R),Kelley1996PRL,Smedberg1999PLB}.
However, due to the effects of the Coulomb barrier \cite{Riisager1992NPA}, the number of proton halo nuclei and candidates is less than that of neutron halo ones.

In 2020, a large mirror asymmetry of $\delta=209(96)\%$ was reported for the transition of $^{22}$Si/$^{22}$O partner to the first $1^+$ excited state of their respective daughter $^{22}$Al/$^{22}$F \cite{Lee2020PRL}.
It was declared in Ref. \cite{Lee2020PRL} that this asymmetry ``reveals the halo structure of $^{22}$Al."
However, neither an enhanced interaction/reaction cross section nor a narrow momentum distribution of breakup fragments was observed for $^{22}$Al.
Moreover, the results presented in Ref. \cite{Lee2020PRL} pertain only to the $1^+$ excited states of $^{22}$Al, while its ground-state configuration remains uncertain.

In this paper, we discuss whether it is solid to suggest $^{22}$Al as a halo nucleus from the existing experimental information.
Besides, based on the state-of-the-art deformed and triaxial relativistic Hartree-Bogoliubov theories in continuum, we examine the possibility of a proton halo in the ground state of $^{22}$Al.

\section{The experiment in 2020}

First, we introduce some details from Ref. \cite{Lee2020PRL} for the convenience of the following discussion.
The asymmetry parameter is defined as $\delta = ft^+ / ft^- - 1$, where $ft^\pm$ is the reduced Gamow-Teller (GT) transition probability of the $\beta^\pm$ decay,
\begin{equation}\label{ft}
ft = \frac{D}{(\frac{g_A}{g_V})^2_{\mathrm{eff}}|M_{\mathrm{GT}}|^2}.
\end{equation}
$D$ and $(g_A/g_V)_{\mathrm{eff}}$ are coupling constants and are considered unchanged for mirror transition processes \cite{Lee2020PRL}.
$M_{\mathrm{GT}}$ is the nuclear matrix element,
\begin{equation}\label{matele}
M_{\mathrm{GT}} = \langle \mathrm{final~state}| \tau\sigma |\mathrm{initial~state}\rangle,
\end{equation}
where $\tau$ and $\sigma$ are the isospin transition and Pauli spin matrices, respectively.
The asymmetry parameter then reads
\begin{equation}\label{delta}
\delta = \frac{ft^+}{ft^- }-1 = \frac{|M_{\mathrm{GT}}^-|^2}{|M_{\mathrm{GT}}^+|^2}-1.
\end{equation}
A basic logic presented in Ref. \cite{Lee2020PRL} can be understood as follows.
\begin{enumerate}[(1)]
    \item If there is a halo in the drip-line nucleus $^{22}$Al, its wave function will be spread out over a wide region.
    \item Its mirror nucleus $^{22}$F is relatively tightly bound with a one-neutron separation energy of 5230(13) keV \cite{AME2020(3)} and is away from the drip line ($^{31}$F \cite{Ahn2019PRL}) by nine neutrons.
    \item The difference in the final states will lead to very different nuclear matrix elements \eqref{matele} for $^{22}$Si $\rightarrow$ $^{22}$Al and $^{22}$O $\rightarrow$ $^{22}$F transitions, giving a large $\delta$ value \eqref{delta}.
\end{enumerate}

While a possible explanation for the observed mirror asymmetry was provided, it should be noted that the above discussion solely concerns the final states in the transitions.
Namely, it is based on a crucial premise that the initial states of $^{22}$Si and $^{22}$O are essentially similar.
This is, however, likely not true.

The $N = 14$ shell closure in $^{22}$O has been recognized for several years \cite{Stanoiu2004PRC,Becheva2006PRL}, and the recent measurement on the rms radii of proton density distributions for oxygen isotopes further supports the double magicity of $^{22}$O \cite{Kaur2022PRL}.
Consequently, $^{22}$O is compact in nature, as also evidenced by its sizable two-neutron separation energy of 10656(57) keV \cite{AME2020(3)}.
In contrast, its mirror partner $^{22}$Si is probably an unbound system beyond the proton drip line.
Based on the observation of $\beta$-delayed two-proton emission in the decay of $^{22}$Si, its two-proton separation energy is deduced to be $-108(125)$ keV \cite{Xu2017PLB}.
In the latest atomic mass evaluation (AME2020) \cite{AME2020(3)}, the extrapolated value for the two-proton separation energy of $^{22}$Si is $-1584(500)$ keV.
Whether $^{22}$Si is unbound or very weakly bound, its density distribution can be rather diffuse, e.g., see results from the relativistic mean-field theory in Ref. \cite{Saxena2017PLB}.
In fact, the two-proton radioactivity of $^{22}$Si and the corresponding decay half-life have been predicted by many theoretical models \cite{Goncalves2017PLB,Sreeja2019EPJA,Cui2020PRC,Xing2021CPC,Mehana2023EPJA}.
Therefore, a significant mirror asymmetry has already existed in the wave functions of $^{22}$Si and $^{22}$O, which would be responsible for the large $\delta$ to a considerable extent.
Given also that the discussion in Ref. \cite{Lee2020PRL} concentrates on the $1^+$ excited state of $^{22}$Al, it warrants careful consideration to conclude that the halo structure of $^{22}$Al is revealed.
Furthermore, even if the $1^+$ state could contain certain $s$ component, it is proton-unbound in nature and, thus, cannot be considered as a halo state.

\begin{figure}[htbp]
  \centering
  \includegraphics[width=0.4\textwidth]{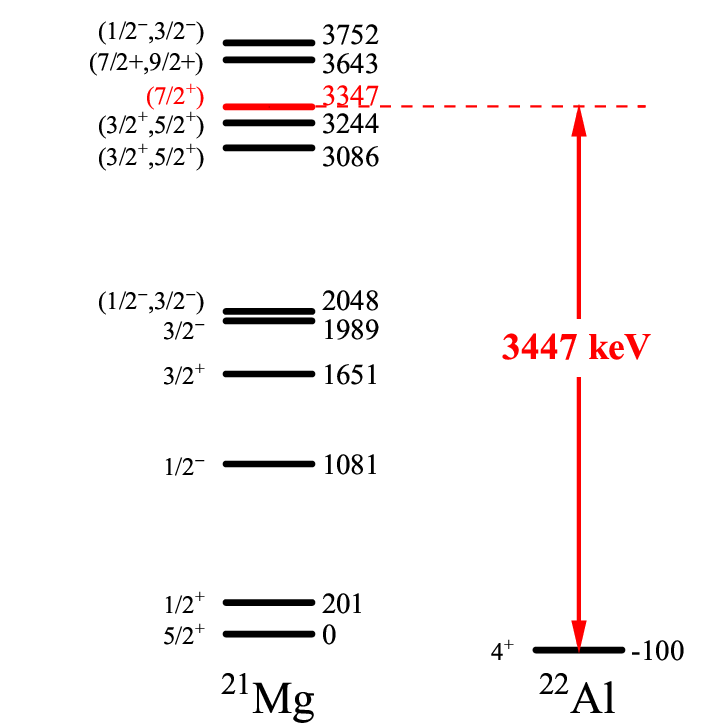}
  \caption{Experimental level schemes for low-lying states below 4 MeV of $^{21}$Mg and the ground state of $^{22}$Al \cite{NNDC,Campbell2024PRL}. Spin parities $J^\pi$ and energies (in keV, with respect to the ground-state $^{21}$Mg, and neglecting uncertainties) are given for each level. The bracket indicates that $J^\pi$ are not completely determined.}
  \label{fig1}
\end{figure}

\section{Ground-state analysis}

Next, we discuss the ground state of $^{22}$Al.
Its mirror nucleus $^{22}$F has a ground-state spin parity of $4^+$ and an excited $3^+$ state at 72 keV.
If the ground-state spin parity of $^{22}$Al is $3^+$, there might be a chance of a halo if it contains enough parentage of $^{21}$Mg (whose ground-state spin parity is $5/2^+$) coupled to a $2s_{1/2}$ proton.
However, the $\beta$ decay of $^{22}$Al strongly suggests a $4^+$ ground-state spin parity \cite{Wu2021PRC}, which necessitates a spin no less than $3/2$ for the valence proton when coupled to the ground-state $^{21}$Mg.
Even if the $4^+$ state has some $\pi2s_{1/2}$ occupation, the parentage has to be to $7/2^+$ or $9/2^+$ states of $^{21}$Mg, and the first candidates are at 3347 and 3643 keV \cite{NNDC}, as shown in Fig. \ref{fig1}.
The one-proton separation energy of $^{22}$Al is recently measured to be 100.4(8) keV \cite{Campbell2024PRL}.
This means that the separation energy of a $2s_{1/2}$ proton from the ground-state $^{22}$Al is at least approximately $3447~(3347 + 100)$ keV, which seems unfavorable for the formation of a proton halo.
Nonetheless, one cannot completely rule out the possibility of a small $s$-orbital component, which might correspond to a definite but not dominant halo \cite{Yang2021PRL}.

In the conventional single-particle shell model, the valence proton in $^{22}$Al would occupy the $1d_{5/2}$ orbital, which could provide the correct spin parity but does not promote the formation of a halo due to the centrifugal barrier.
Since $^{22}$Al is an open-shell drip-line nucleus, the deformation, pairing correlations, and continuum effects play indispensable roles in its microscopic description.
Therefore, the deformed relativistic Hartree-Bogoliubov theory in continuum (DRHBc) \cite{Zhou2010PRC(R),Li2012PRC,Li2012CPL,Zhang2020PRC} could provide valuable insights.
The DRHBc theory has been successfully applied to the microscopic description of known halo nuclei $^{17,19}$B \cite{Yang2021PRL,Sun2021PRC(1)}, $^{15,19,22}$C \cite{Sun2018PLB,Sun2020NPA}, $^{31}$Ne \cite{Zhong2022SciChina,Pan2024PLB}, and $^{37}$Mg \cite{Zhang2023PLB,An2024PLB}, the prediction of new halo candidates $^{39}$Na \cite{Zhang2023PRC(L1)} and $^{42,44}$Mg \cite{Zhou2010PRC(R),Li2012PRC,Zhang2019PRC}, and the exploration of the limits of nuclear existence \cite{In2021IJMPE,Zhang2021PRC(L),Pan2021PRC,He2021CPC,He2024PRC,Zhang2022ADNDT,Guo2024ADNDT}.
While mean-field calculations cannot yield states with well-defined spins, in the DRHBc theory the angular momentum can be analyzed based on the main components of the valence nucleon(s) and compared with experiments, as a spherical Dirac Woods-Saxon basis \cite{Zhou2003PRC,Zhang2022PRC} that possesses good quantum numbers is used to expand wave functions.
In this regard, the implementation of angular momentum projection in the DRHBc theory \cite{Sun2021SciBull,Sun2021PRC(2)} for odd-odd nuclei in future work is also desirable.

In Ref. \cite{Pan2022PRC}, the potential energy curves with different proton and neutron orbitals blocked have been constructed for $^{22}$Al using the DRHBc theory with the density functional PC-PK1 \cite{Zhao2010PRC}.
It can be found in Figs. 1 and 2 of Ref. \cite{Pan2022PRC} that the ground state of $^{22}$Al is deformed with $\beta_2 \approx 0.3$, and the valence proton occupies an orbital with quantum numbers $\Omega^\pi = 5/2^+$, where $\pi$ denotes the parity and $\Omega$ is the projection of the total angular momentum on the symmetry axis.
The results show that this orbital has more than $99\%$ $1d_{5/2}$ component.
The $1d_{5/2}$ component is also overwhelmingly dominant for the valence neutron in both $^{22}$Al and $^{21}$Mg.
This is consistent with the $5/2^+$ spin parity for the ground-state $^{21}$Mg and the fact that the $4^+$ ground state of $^{22}$Al can be obtained from the parentage of the ground state of $^{21}$Mg coupled to a $1d_{5/2}$ proton.
However, we note that the $2s_{1/2}$ component cannot contribute to the valence proton in $^{22}$Al since its total angular momentum quantum number $j = 1/2$ is smaller than $\Omega=5/2$.
Therefore, the ground-state halo structure of $^{22}$Al is not supported by the DRHBc theory that assumes the axial symmetry.

A recent calculation based on a Woods-Saxon potential suggests that the region of halo nuclei could be extended by triaxial deformation that allows the appearance of $s$- or $p$-wave components in some weakly bound orbitals \cite{Uzawa2021PRC(L)}.
It would be interesting to examine whether triaxial deformation plays a role in the ground state of $^{22}$Al.
To this end, we resort to the triaxial relativistic Hartree-Bogoliubov theory in continuum (TRHBc) \cite{Zhang2023PRC(L2)}, which incorporates triaxiality, pairing correlations, and continuum effects in a microscopic and self-consistent way.

The calculated one-proton separation energy is positive for $^{22}$Al but negative for $^{21}$Al, i.e., the proton drip-line location for aluminum isotopes \cite{Wu2021PRC} is correctly reproduced by the TRHBc theory \cite{Zhang2023PRC(L2)}.
The TRHBc results from various density functionals are also in good agreement with available data for neutron separation energies and charge radii of aluminum isotopes \cite{Zhang2023PRC(L2)}.
Here, taking the density functional PC-F1 \cite{Burvenich2002PRC} as an example, we focus on the ground-state properties of $^{22}$Al.
Other numerical details in our calculation are the same as those used in Ref. \cite{Zhang2023PRC(L2)}.

\begin{figure}[htbp]
    \centering
    \includegraphics[width=0.5\textwidth]{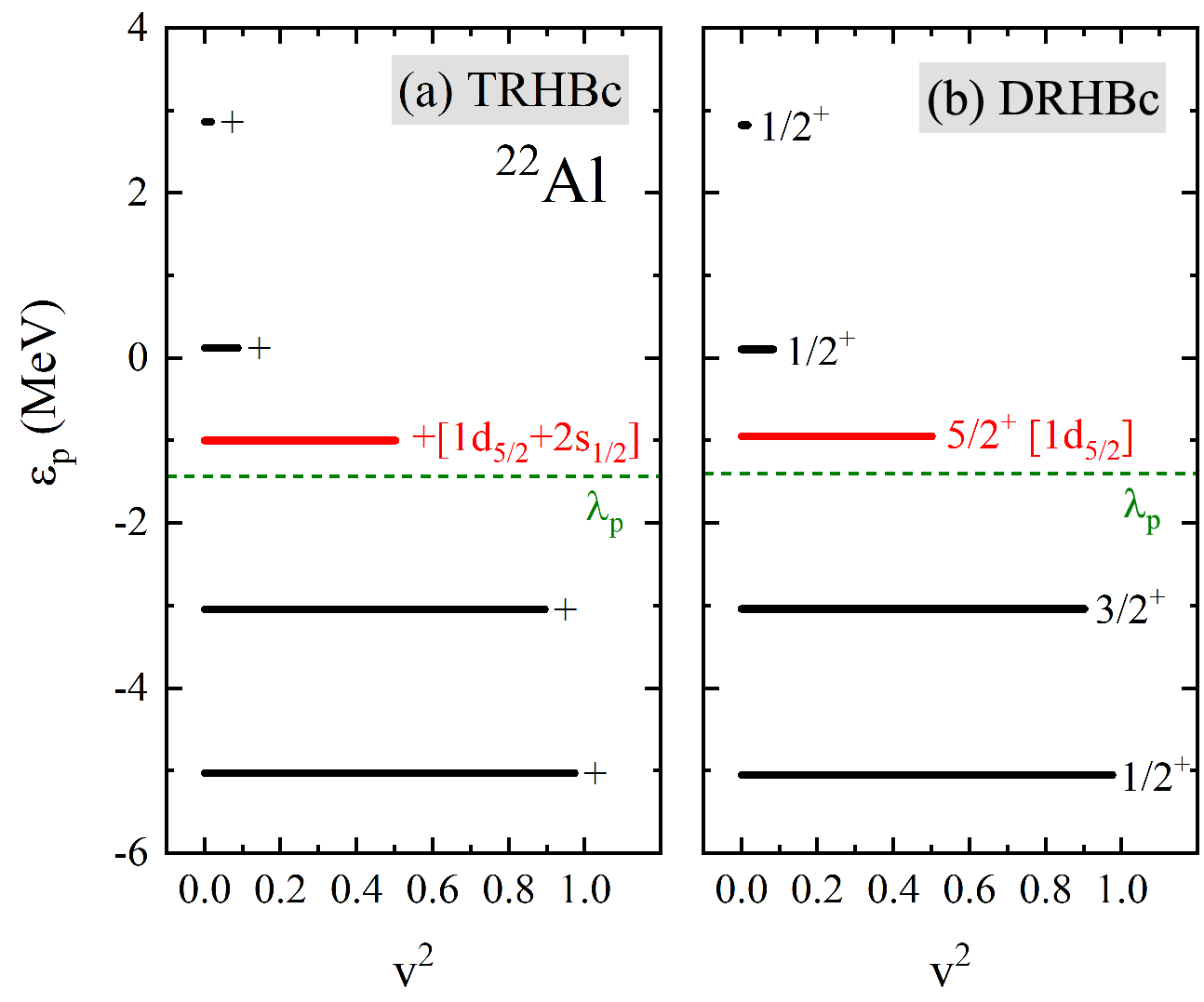}
    \caption{Energy $\varepsilon_p$ versus occupation probability $v^2$ for single-proton orbitals around the Fermi energy $\lambda_p$ (dashed line) in the canonical basis for $^{22}$Al from (a) TRHBc and (b) DRHBc calculations. Each orbital is labeled by quantum number(s) $\pi$ in (a) and $\Omega^\pi$ in (b). The orbital occupied by the valence proton is in red, and its components are given in the square brackets.}
    \label{fig2}
\end{figure}

The calculated binding energy of $149544.8$ keV reproduces well the recently measured result of $149313.1(3)$ keV \cite{Campbell2024PRL}, with a discrepancy below $0.01\%$.
However, the calculated one-proton separation energy of $21.7$ keV is only about one-fifth of the experimental one, $100.4(8)$ keV \cite{Campbell2024PRL}, which means that the TRHBc theory describes $^{22}$Al as more weakly bound than it actually is.
The ground-state deformation parameters are $\beta = 0.30$ and $\gamma = 7.93^\circ$ for $^{22}$Al, that is, slight triaxial deformation occurs.
To explore the effects of triaxial deformation, we exhibit in Fig. \ref{fig2}(a) the single-proton orbitals around the Fermi energy in the canonical basis for $^{22}$Al from the TRHBc calculation, in comparison with those from the DRHBc calculation in Fig. \ref{fig2}(b).
We find that the influence of triaxial deformation is on the order of 0.01 MeV for both the single-proton energies and the binding energy.
Nonetheless, due to the breaking of axial symmetry, $\Omega$ is no longer a good quantum number,
which allows the appearance of $2s_{1/2}$ component in the weakly bound orbital occupied by the valence proton.
This is similar to the scenario discussed in Ref. \cite{Uzawa2021PRC(L)} for nuclei with $N=13$ based on a Woods-Saxon potential; here, $Z=13$.
Quantitatively, there are about $93\%$ $1d_{5/2}$ and $5\%$ $2s_{1/2}$ components.

\begin{figure}[htbp]
  \centering
  \includegraphics[width=0.45\textwidth]{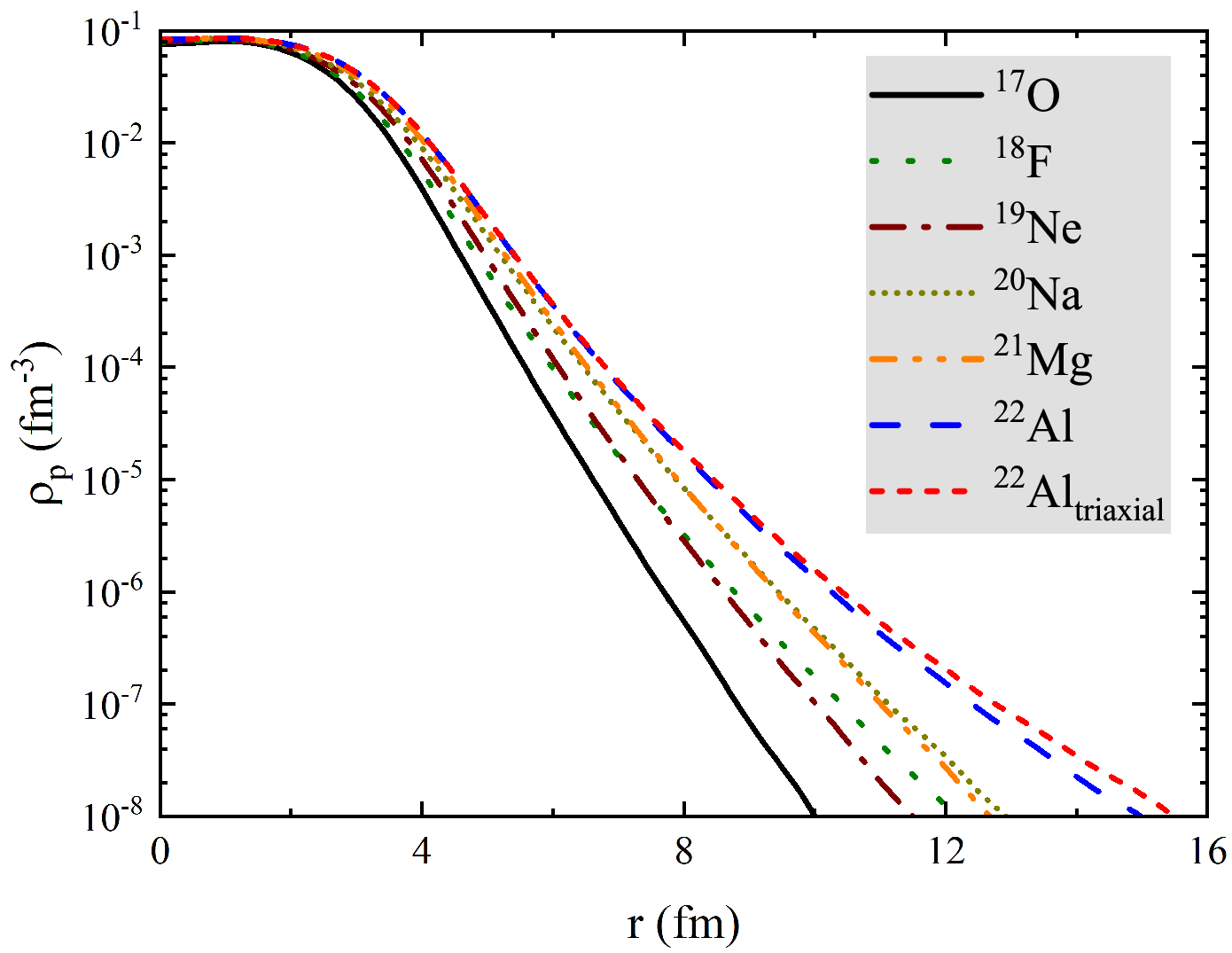}
  \caption{Angle-averaged proton density as a function of radial coordinate $r$ for $N =9$ isotones with the proton number ranging from 8 ($^{17}$O) to 13 ($^{22}$Al). For $^{22}$Al, the TRHBc and DRHBc results are shown by short-dashed and dashed lines, respectively. For other isotones, the TRHBc and DRHBc theories yield the same results.}
  \label{fig3}
\end{figure}

The inclusion of $2s_{1/2}$ component in the valence proton orbital is expected to extend the proton density distribution in $^{22}$Al.
To investigate this effect, we present the angle-averaged proton densities for $N =9$ isotones with $8\le Z\le 13$ in Fig. \ref{fig3}.
The TRHBc and DRHBc results for $^{22}$Al are represented by short-dashed and dashed lines, respectively, while for other lighter isotones the TRHBc and DRHBc results are identical.
Specifically, $^{17}$O, $^{18}$F, $^{19}$Ne, $^{20}$Na, and $^{21}$Mg are predicted to be axially deformed in their ground states with $\beta_2 = -0.08$, $0.27$, $0.26$, $0.42$, and $0.32$, respectively.
Figure \ref{fig3} shows that near the central region ($r\lesssim 4$ fm), the proton density gradually increases with $Z$.
At larger $r$, there is a notable increase in the proton density from an even-$Z$ isotone to the next heavier odd-$Z$ isotone.
The proton density of this odd-$Z$ isotone can even exceed that of the next heavier even-$Z$ isotone beyond $r \approx 10$ fm.
These are primarily due to the blocking effect of the odd nucleon in extending density \cite{Sun2020NPA}, with deformation effects also contributing.
For $^{22}$Al, the difference between the TRHBc and DRHBc results becomes evident only in the region of $r\gtrsim 10$ fm.
In this region, the proton density of $^{22}$Al with $Z=13$ is approximately one order of magnitude lower than the neutron density of the known neutron halo nucleus $^{17}$B with $N =12$ (see Fig. 3 of the supplemental material of Ref. \cite{Yang2021PRL}).

\begin{figure}[htbp]
  \centering
  \includegraphics[width=0.45\textwidth]{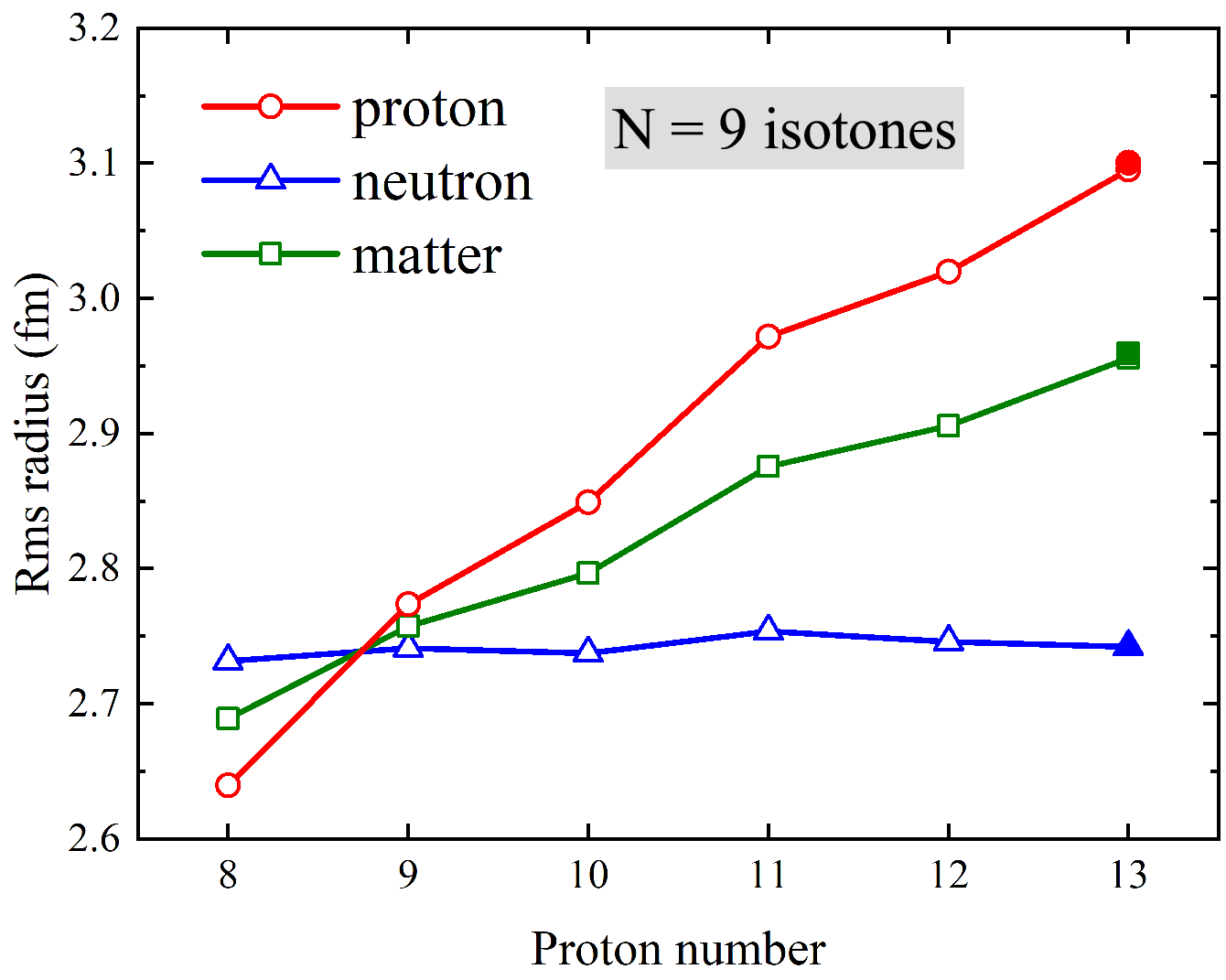}
  \caption{Proton, neutron, and matter rms radii for $N =9$ isotones with the proton number ranging from 8 ($^{17}$O) to 13 ($^{22}$Al). For $^{22}$Al, the TRHBc and DRHBc results are shown by solid and open symbols (almost overlapped), respectively. For other isotones, the TRHBc and DRHBc theories yield the same results that are shown by open symbols.}
  \label{fig4}
\end{figure}

The slightly more diffuse proton density, attributable to the minor $2s_{1/2}$ component, corresponds to a marginal increase in the proton rms radius of $^{22}$Al.
This is depicted in Fig. \ref{fig4}, which presents the proton, neutron, and matter rms radii for $N=9$ isotones with $8 \le Z \le 13$.
While the proton rms radius increases steadily with $Z$, the neutron rms radius remains almost unchanged, and the matter rms radius lies between the two.
The differences in neutron and matter rms radii between the TRHBc and DRHBc results for $^{22}$Al are almost indistinguishable.
As shown in Fig. \ref{fig4}, the most significant increase in the proton rms radius occurs from $^{19}$Ne to $^{20}$Na, due to deformation effects, rather than from $^{21}$Mg to $^{22}$Al.
Consequently, the $2s_{1/2}$ component of only $5\%$ appears insufficient to form a discernible proton halo in the present calculation.

Of course, the $s$-wave component may vary depending on the employed pairing strength and density functional.
Further calculations indicate that triaxial deformation and the $2s_{1/2}$ component can be enhanced by pairing.
For instance, increasing the pairing strength from $V_0 = -342.5$ MeV fm$^3$ to $-350$ MeV fm$^3$ results in deformation parameters for $^{22}$Al of $\beta =0.29$ and $\gamma =9.33^\circ$, with the $2s_{1/2}$ component increasing to $8\%$.
However, this enhanced pairing renders the one-proton separation energy of $^{22}$Al negative, as $^{22}$Al gains less binding energy than the even-$Z$ nucleus $^{21}$Mg.
While the TRHBc calculations with other density functionals such as PC-PK1 yield similar results as above, the \textit{ab initio} valence-space in-medium similarity renormalization group calculations employing four sets of chiral interactions predict a somewhat larger $\pi2s_{1/2}$ occupation of $\lesssim 0.2$ for the ground state $^{22}$Al \cite{Sun2024CPC}.
Additionally, a larger $\pi2s_{1/2}$ occupation of 0.29 is predicted by shell-model calculations using USD Hamiltonians \cite{Campbell2024PRL}.

It was generally accepted that a significant quantity of low-$\ell$ components are necessary for halo formation.
However, a small $s$-wave component of $9(2)\%$ has been observed and reproduced by the DRHBc theory for the halo nucleus $^{17}$B, indicating that the dominant occupation of $s$ or $p$ orbitals is not a prerequisite for a neutron halo \cite{Yang2021PRL}.
On the other hand, the Coulomb barrier hinders the formation of a proton halo.
Therefore, here we cannot draw a firm conclusion on whether $^{22}$Al is a halo nucleus,
but our results demonstrate that the appearance of $s$-wave components for the valence proton in $^{22}$Al necessitates the consideration of triaxial deformation at the mean-field level.

\section{Summary}

In summary, we argue that it is unconvincing to suggest a proton halo in $^{22}$Al based on the large mirror asymmetry observed in the $^{22}$Si/$^{22}$O $\rightarrow$ $^{22}$Al/$^{22}$F transitions.
First, this asymmetry can be largely attributed to the differences between the wave functions of the doubly magic nucleus $^{22}$O and the likely unbound nucleus $^{22}$Si.
Second, the transitions involve the $1^+$ excited states rather than the ground states of $^{22}$Al and $^{22}$F.
Third, even if the $1^+$ state of $^{22}$Al contains some $s$-wave component, it cannot be considered a halo state due to its proton-unbound nature.

An analysis of the spin parity suggests that a weakly bound $2s_{1/2}$ valence proton in the ground-state $^{22}$Al is improbable.
The ground state of $^{22}$Al is investigated using the DRHBc and TRHBc theories.
A small $s$-wave component of $5\%$ appears for the weakly bound valence proton in $^{22}$Al only when triaxial deformation is considered.
Examination of densities and rms radii indicates that this small $s$-wave component is insufficient to form a discernible proton halo in $^{22}$Al.
Nevertheless, other recent theoretical calculations have yielded slightly larger $\pi2s_{1/2}$ occupations for the ground state of $^{22}$Al.
Since how many low-$\ell$ components are sufficient to form a proton halo in the presence of the Coulomb barrier remains an open question, a firm conclusion on whether $^{22}$Al is a halo nucleus has not been obtained.
Future measurements of reaction or interaction cross sections and momentum distributions of breakup fragments for $^{22}$Al are highly desirable.

\begin{acknowledgments}
We greatly appreciate the anonymous referees for extending our discussion to involve the ground-state spin parity, densities, and rms radii.
We thank X. Lu, J. Meng, P. Papakonstantinou, and S. Q. Zhang for helpful suggestions and discussions.
This work was partly supported by the National Natural Science Foundation of China (Grants No. 12305125 and No. 12205030) and the Sichuan Science and Technology Program (Grant No. 2024NSFSC1356).
\end{acknowledgments}


%

\end{document}